\documentclass{nature_withfigure}
\usepackage{graphicx}
\usepackage{amsmath,amsthm,amssymb}
\usepackage{amssymb}
\usepackage{bm}
\usepackage[labelfont=up]{caption}
\usepackage{hyperref}
\usepackage{lineno}
\usepackage{footmisc}
\usepackage[usenames, dvipsnames]{color}

\def\apj{Astrophysical Journal}                
\def\mnras{Monthly Notes of Royal Astronomical Society}
\def\apjl{Astrophysical Journal, Letters}

\def\araa{Annual Review of Astronomy \& Astrophysics}

\def\prd{Phyical Review D}

 \title{Reinterpreting Low Frequency LIGO/Virgo Events as Magnified Stellar-Mass Black Holes at Cosmological Distances.}

\author{Tom Broadhurst$^{1,2}$, Jose M. Diego$^{3}$, George Smoot III $^{4, 5,6,7}$}

\begin{document}

\maketitle

\begin{affiliations}

\item {Department of Theoretical Physics, University of Basque Country UPV/EHU, E-48080 Bilbao, Spain}
\item {IKERBASQUE, Basque Foundation for Science, E-48013 Bilbao, Spain}
\item {Instituto de F\'isica de Cantabria, CSIC-Universidad de Cantabria, E-39005 Santander, Spain}
\item {IAS TT \& WF Chao Foundation Professor, IAS, Hong Kong University of Science and Technology,
Clear Water Bay, Kowloon, 999077 Hong Kong} 
\item {Paris Centre for Cosmological Physics, APC, AstroParticule et Cosmologie, Universit\'{e} Paris Diderot,
CNRS/IN2P3, CEA/lrfu, 
Universit\'{e} Sorbonne Paris Cit\'{e}, 10, rue Alice Domon et Leonie Duquet,
75205 Paris CEDEX 13, France}  
\item {Physics Department and Lawrence Berkeley National Laboratory, University of California, Berkeley, 94720 CA, USA}
\item {Physics Department and Energetic Cosmos Laboratory, Nazarbayev University, Astana, Kazakhstan}
\end{affiliations}

\begin{abstract}
\boldmath

Gravitational waves can be focused by the gravity of an intervening galaxy, just like light, thereby magnifying binary merging events in the far Universe\cite{Wang}. High magnification by galaxies is found to be responsible for the brightest sources detected in sky surveys\cite{Wardlow,Negrello,Bussmann}, but the low angular resolution of LIGO/Virgo is insufficient to check this lensing possibility directly. Here we find that the first six binary black hole (BBH) merging events reported by LIGO/Virgo\cite{Abbot2} show clear evidence for lensing in the plane of observed mass and source distance. The four highest mass events follow an apparent locus in this plane, which we can reproduce by galaxy lensing, where the higher the magnification, the generally more distant the source so the wave train is stretched more by the Universal expansion, by factors of $2-4$. This revises the reported BBH distances upwards by an order of magnitude, equal to the square root of the magnification. Furthermore, the reported black hole masses must be decreased by $2-4$ to counter the larger stretch factor, since the orbital frequency is used to determine the black hole masses. This lowers the masses to $5-15M_\odot$, well below the puzzlingly high values of $20-35M_\odot$ otherwise estimated\cite{Abbot1}, with the attraction of finding agreement in mass with black holes orbiting stars in our own Galaxy\cite{Remillard}, thereby implying a stellar origin for the low frequency events in the far Universe. We also show that the other two BBH events of higher frequency detected by LIGO/Virgo, lie well below the lensing locus, consistent with being nearby and unlensed. If this apparent division between local and distant lensed events is reinforced by additional new detections then the spins and masses of stellar black holes can be compared over a timespan of 10 billion years by LIGO/Virgo.

\end{abstract}



Prior to the first LIGO detection\cite{Abbot1}, merging binary black holes (BBH) chirp masses of $5-15M_\odot$ were anticipated, bracketing the range of the established black holes seen orbiting stars within our Galaxy, and consistent with prior predictions of binary star evolution of solar metallicity\cite{Dominik}. \footnote{The chirp mass of a BBH is derived from the combination of the two individual masses in such a way that when the individual masses are comparable, the chirp mass is also comparable but when one of the masses is much lighter, the chirp mass of the BBH is closer to the lighter mass component.} 
Instead, four of the six BBH events now detected by LIGO/Virgo have black hole masses reported in the range $20-35M_\odot$, that may arise preferentially from metal poor stellar binary progenitors \cite{Belczynski} with minimal mass loss by stellar winds \cite{Abbot3}. Since such binaries are hypothetical, speculation regarding primordial black holes of $30M_\odot $, has been revived that can qualify as cold dark matter\cite{Bird,Carr}. However, this interpretation is directly unsupported by the observed high degree of microlensing observed through large dark matter columns in galaxy clusters\cite{Kelly,Venumadhav,Diego,Oguri}, and is disfavoured dynamically\cite{Brandt,Koushiappas,Green} and by a lack of accreting gas\cite{Gaggero,Ali-Hamoud}.

 Lensing offers a solution with lower black hole masses and substantially higher redshifts. Firstly it is important to appreciate that the redshift of BBH events cannot be derived directly from the inspiral wave frequencies because cosmological stretching of the wave train by $(1+z)$ is exactly compensated by reducing the source frame chirp mass, $M_{chirp}$, by the same factor\cite{Finn}. In terms of the observed ``chirp mass", $M_{\rm Chirp}(z)$, in the detector frame this means:
\begin{equation}
 M_{\rm Chirp}(z) = (1+z) M_{\rm Chirp} = (1+z) \frac{(m_1 m_2)^{3/5} }{ (m_1 +m_2)^{1/5}}
\end{equation}
where $m_1$, and $m_2$ are rest frame masses, so: $M_{\rm Chirp}(z)=(1+z)M_{\rm Chirp}(0)$. Because gravitation is scale free, GWs from a local binary with masses $(m_1, m_2)$ are indistinguishable by frequency dependence from distant masses $(\frac{m_1}{1 + z} , \frac{m_2}{ 1 + z})$. The cosmological time dilation ensures that the waveforms are shifted by precisely the necessary amount to preserve the chirp frequency profile. 

The distances published for the LIGO/Virgo events are instead obtained by comparing the observed GW signal strength to the intrinsic strength predicted by General Relativity as a function of $M_{\rm Chirp}(z)$ ($m_1,m_2$), where the measured strain 
$h(t)$, is inversely proportional to the luminosity distance $D_L$ allowing an indirect estimate of the 
redshift, $z_{ind}$, to be inferred:
\begin{equation}
D_L(z_{ind}) =\frac{{{\rm M}_{chirp}^{5/6}(z)}}{{h(t))}} F(t,{\rm M}_{obs},\Theta) = \frac{d_L}{\sqrt{\mu}}
\label{Eq_ht}
\end{equation}
where $F(t,{\rm M}_{obs},\Theta)$ combines the angular sky sensitivity, orbital inclination, spin and polarization of the binary source and its distribution is numerically estimated with a $\simeq 40\%$ dispersion\cite{Abbot1,Ng}. The co-alignment of the LIGO interferometers means the polarization is uncertain, resulting in large errors in the inferred luminosity distance, $D_L$ (Figure~1)\cite{Abbot1}, but this will improve for sources detectable by more interferometers.
A gravitational wave source at true luminosity distance $d_L$ which is gravitationally lensed with magnification will $\mu$ has its observed strain $h(t)$ a factor of $\sqrt{\mu}$ larger and the source will be inferred to be that much closer.  The possibility of magnification, $\mu$, has not been considered in the published mass and distance estimates, with the implicit assumption made that $\mu=1$, perhaps because lensing is rare for a randomly sampled event with optical depths of only $\approx 5\times10^{-5}$ for $\mu=10$ at $z\approx 1$ calculated for strongly lensed sources magnified by intervening galaxies\cite{Turner}\footnote{Our lensing kernel predicts a slightly larger value of $7\times10^{-5}$ for $\mu=10$ at $z=1$}. However, BBH events can currently be detected by LIGO/Virgo to only modest redshift, $z \simeq 0.15$, for $10M_\odot$, whereas gravitational lensing accesses a much larger volume. This is particularly true for compact sources which can by highly magnified near lensing caustics, permitting detection to $z \simeq 2$ (as we show below), thereby compensating for the low optical depth with an enhanced volume ratio of $\simeq 1000$. This is further enhanced by BBH evolution linked to cosmic star formation, which peaks at $z \simeq 2$ with an order of magnitude or more rate enhancement\cite{Madau}, so a significant presence of lensed events should not be surprising.

 The lensing enhancement of the GW signal strength, $h(t) \propto \sqrt{\mu}$, means the inferred luminosity distance, $D_L(inferred)$ underestimates the real source distance, $d_L$ by the square root of the magnification\cite{Dai1,Ng}: $D_L=d_L/\sqrt{\mu}$. For example, a lensed source at $z_s=2.0$ will be assigned a low redshift of $z_{inferred}=0.15$ when magnified by a factor of $\simeq 200$, and hence an intrinsic chirp mass of $10M_\odot$ is overestimated by a factor of $3.0/1.15$, becoming $M_{chirp}(z)\simeq 30M_{\odot}$, similar to the puzzlingly high chirp mass events of low frequency detected by LIGO/Virgo. Such high magnifications are not unusual for compact like AGN\cite{Broadhurst} and lensed high redshift stars discovered recently, for which magnifications exceeding $10^{3}$ are found\cite{Kelly,Diego,Oguri}. The region of intense GW emission detected by LIGO/Virgo is tiny, $ <10^3$km, and hence this can lie very close in projection to a lensing caustic, receiving high magnification. 

In Figure~1 we plot the chirp masses and distances inferred by LIGO/Virgo assuming no lens magnification for the six BBH detections, with a surprising result that the four lowest frequency (high mass) events follow a surprising trend of increasing chirp mass with decreasing distance, whereas the two relatively low mass events of $M_{chirp}=8-9M_\odot$ lie well off this apparent locus, falling in the mass range of black holes in our Galaxy. This behaviour is more puzzling when considering just how much further away the high chirp mass events could have been detected given their measured signal strength. In particular, the two largest chirp mass events lie within just the closest 8\% and 16\% of their maximum available volumes. The opposite behaviour should be expected given the expected strong evolution, so that most events should lie at the upper end of their detectable volumes, near the detection threshold, with a mean $V/V_{max}\simeq 0.7$\cite{Schmidt}.  

  We now populate Figure~1 with events predicted by lensing due to the intervening galaxy population, together with the distribution of unlensed events for various illustrative choices of BBH mass function and event rate evolution. The prediction follows the well understood optical depth that is dominated by galaxy lensing (not clusters\cite{Smith}) that accounts for the recent Herschel satellite detection of large samples of lensed high redshift star forming galaxies\cite{Wardlow,Negrello,Bussmann}. Micro-lensing can be neglected as the  wavelengths emitted are so long that stars within galaxies are too small to act as deflectors\cite{Nakamura}. The form of the high magnification relation, $P(>\mu)\propto \mu^{-2}$, is strictly set by the universal caustic relation, and has been found to be precisely followed with ray tracing through large cosmological simulations\cite{Hilbert,Zackrisson}. We include the frequency sensitivity of LIGO which rapidly raises at low frequency making chirp masses higher than $\simeq 35M_\odot$ hard to detect.

 This well defined calculation of optical depth as a function of magnification is now applied to the BBH source population for illustrative choices of mass function and event rate evolution. Irrespective of the details we see in Figures~1 and 2 that some proportion of lensed events is inevitable at higher chirp mass, with the most magnified events misattributed to the smallest inferred distances. A mildly log-normal mass function matches the data well, as shown in Figure~1 (left), motivated by the Galactic mass distribution, centered on $\simeq 8M_\odot$, with a modest dispersion in the log of $\Delta\log M_{Chirp}=1.5M_\odot$ (Figure~3), and normalized at the lower end of published estimates of $\simeq 10\, {\rm events}/yr/Gpc^3$ (Figure~3), reflecting just the two low chirp mass events, for self consistency, as these have masses falling squarely within the mass function used here. We can understand the observed trend of increasing chirp mass with decreasing distance follows from a generally increasing trend of magnification with increasing redshift, as seen in Figure~1 (right), with redshift in the range $0.5<z<3.0$ and intrinsic chirp masses reduced by $(1+z)$. For this solution the two low mass events are well centered within the predicted distribution of unlensed events, as shown in Figure~1 (left). Relatively steeper evolution (Figure~3) helps enhance the proportion of lensed events to match the data and the redshift separation between the unlensed and lensed events produces a central "gap" in the the chirp mass distribution. This evolution is more in line with binary star coalescence, which is predicted to decline by nearly two orders of magnitude since $z\simeq 2-3$, favouring sub-solar metallicity\cite{Dominik} and for which BBH formation is predicted to be relatively efficient\cite{Belczynski}.

For narrower choices of intrinsic mass functions (Figure~3), using a Gaussian with $\sigma=1.5M_\odot$ or a delta function centered at $8M_\odot$ (Figure~3), the predicted lensed sources fall short of the two most massive LIGO/Virgo events, as shown in Figure~2 (left \& center), indicating we need a modest tail of high mass events, as in Figure~1, or perhaps black holes masses that were somewhat higher at $z>1$ than today. This interesting possibility of mass evolution can be tested in a model independent way with more detections, because the lensing predictions can be based directly on the observed distribution of low chirp mass events.The lowest proportion of lensed events is produced by a broad log-normal mass function, Figure~2 (right), as then unlensed events dominate because the relatively massive tail is detectable to greater depth, and so lensed events become a small proportion, confirming earlier work based on relatively shallow power-law mass function\cite{Dai1,Ng}. But this solution over produces low chirp mass events with a distribution quite unlike the current data, which has relatively few low mass events, Figure~2 (right).
  
At high magnifications, multiply lensed events are expected with time delays of hours-week particularly for events with $M_{chirp} > 15M_\odot$ where lensing appears to dominate. The most magnified sources near the cusp of a lensing caustic will lie predominantly in the relatively large area outside caustic cusps, resulting in a pair of images with a large magnification ratio\cite{Broadhurst} where the weaker event appears last due to the longer time delay through the center of the potential\cite{Blandford} and may suffer a distinguishing phase shift\cite{Dai2}, but generally will fall well below the current detection threshold. Detectable repeated events are currently restricted to those of similar strength, within a factor of two in magnification, as currently the detections span only a factor of two in signal-to-noise. This includes image pairs closely straddling the critical curve, with very similar high magnifications, where a time delay of a few hours can be expected. The most overhead event will provide the strongest detection, with the other event usually below the detection limit, as the detectors are revolved by Earths rotation away from the overhead position, falling to zero at the horizon. Such events may be recovered from the data stream at lower SNR and should be found more easily with additional interferometers, covering a wider longitude range with more uniform sensitivity.

In conclusion, it would be surprising if the strain limited detections from LIGO/Virgo did not contain a significant proportion of highly gravitationally lensed events from higher redshifts, beyond the nominal unlensed detection volume. If the source of BBH has a similar mass distribution to the Galactic binary BHs, then these highly lensed events would show up as having apparently higher masses because of the cosmological redshift of the chirp to lower frequency. If the parent BH mass distribution is narrow and there is strong evolution of such binaries, then half the detected events might be the result of strong magnification from higher redshift. This is what we claim is most consistent with the observed LIGO/Virgo events detected thus far. We can predict then that future observations will see a similar distribution and that it is possible that at some point two images of the same event might be detected as more detectors come on line. We also predict that a reduction in low frequency detector noise will result in the detection of events with very high chirp masses, up to $\simeq 50M\odot$, corresponding to lensed sources in the high mass tail, $M_{chirp}>10M_\odot$ with $z>2$.



\subsection{} {T.J.B. thanks the IAS/HKUST for generous hospitality. The authors wish to thank KS Cheng, Nick Kaiser, Tjonnie Li, Ken Ng, Henry Tye and K. Belczynski for useful conversations. J. M. D. acknowledges the support of project AYA2015-64508-P(MINECO/FEDER, UE) and hospitality of UPenn.

\subsection{}{All authors contributed significantly to this work.}

\subsection{Author Information} The authors declare no competing financial interests. Correspondence and requests for materials should be addressed to tom.j.broadhurst@gmail.com

\subsection{Competing financial interests} The authors declare no competing financial interests.

\begin{methods}
In this section we describe the calculations made to compute the probability of lensing and the MonteCarlo simulation of gravitationally lensed GW. 
\subsection{Estimating the lensing probability:}
For the gravitational lenses, we assume the mass function for haloes of \cite{Watson13}. This mass function is derived from a suite of N-body simulations at different resolutions including the JUBILEE simulation \cite{Watson14}, one of the largest N-body simulations ever made with $\approx 630$ Gpc$^3$  volume (for $h=0.7$). The mass function covers a wide mass range and improves on other popular mass functions in the high end of the mass owing to the uniquely large volume of JUBILEE. 
For the computation of the lensing probability we restrict ourselves to haloes between $10^{12} {\rm M}_{\odot}$ and $3\times10^{15} {\rm M}_{\odot}$ which is the relevant mass range contributing to the lensing probability, specially for large magnification factors (see Figure~4 in \cite{Hilbert} for instance).
For a given halo, we assume an elliptical NFW profile with ellipticity $e=0.2$, virial radius scaling as $M_{vir}^{1/3}$ and concentration given by the model in \cite{Prada}. Through inverse ray tracing we compute the lensing probability (in the source plane) for different combinations of the redshift of the lens and redshift of the source, $p(>\mu,z_s,z_l,M_l)$. This computation is repeated for different halo masses, halo redshift and source redshift.  The global probability of lensing of a source at redshift $z_s$  is computed after integrating the lensing probabilities of all individual haloes in the mass range considered and up to the redshift of the source. 
\begin{equation}
P(>\mu,z_s)=\int_0^{z_max}dV(z_l)dz_l\int_{M1}^{M2}dM_l \frac{dN}{dMdz}p(>\mu,z_s,z_l,M_l)
\end{equation}
where $dN/dMdz$ is the halo mass function and $dV(z_l)$ is the volume element at redshift $z_l$. 
We set a maximum redshift for the lenses of $z_{max}=2$. As discussed earlier, we set $M1$ and $M2$ to $10^{12}$ and $3\times10^{15}$ solar masses respectively. The main contribution to the probability is obtained below redshift $z_l=1$ and declines sharply beyond $z_l=1$ with negligible contributions beyond $z_l=2$. 
Figure~\ref{fig:result4} shows the probability of having a background source at $z_s=2$ magnified by a factor larger than $\mu$. The dotted line shows a model falling like $\mu^{-2}$ which is the expected behaviour of the lensing probability near a fold caustic at high magnification factors. Beyond $\mu\approx80$ the lensing probability follows closely this theoretical model. The apparent plateau in the probability at intermediate magnifications ($\mu\approx 20--60$) is present in all the individual haloes that are supercritical and emerges from the discontinuity in the magnification between the interior and exterior sides of the caustic region. 

From Fig.~\ref{fig:result4}, at $\mu>100$ the probability is $\approx 3\times10^{-7}$. A redshift slice between $z=1.8$ and $z=2.2$ contains approximately 200 Gpc$^3$. If at $z\sim 2$ the rate  is  $\approx 3\times10^4$ events per year and per Gpc$^3$ (see Fig.~\ref{fig:result3}), the redshift slice $1.8<z<2.2$ will contain $\approx 6\times10^6$ events per year. That is, we would expect $\approx 2$ events per year with magnification larger than 100 from that redshift slice. If instead one considers a much more conservative rate at $z \sim 2$ of $\approx 10^3$ merging events per year, per Gpc$^3$, then we would expect 1 lensed event per $\approx 15$ year interval with $\mu>100$ and between $z=1.8$ and $z=2.2$. 

In general, since the probability of having $\mu>1$ is significantly smaller than 1, the fraction of events that are lensed at a given redshift is much smaller than the number of events that are not lensed at the same redshift. However, since through large magnifications we can access a much larger volume, this reduction in the probability is in part  compensated by the increase in volume. If in addition, the intrinsic rate of events at higher redshift is larger, this boosts the probability of observing a lensed event. 

\subsection{Estimating rates. MonteCarlo simulation:}
Given a rate for events at a redshift $z_s$ and a distribution function for the intrinsic chirp masses (see Fig.~\ref{fig:result3}), we compute the number of events at $z_s$ that would be observed above the detection limit of LIGO. We set a conservative limit of $SNR>10$ above which most LIGO events should have been detected. The number of observed events depends significantly on the distribution of geometric factors, $\Theta$, (see Eq.~\ref{Eq_ht}) for which we follow \cite{Finn} and use 
\begin{equation}
P(\Theta)= \frac{5\Theta(4-\Theta)^3}{256}
\end{equation}
where $\Theta$ takes values between 0 and 4.  The number of observed and not lensed events (i.e assuming $\mu=1$) is computed after integrating the mass function of BBH, the rate density of mergers and the geometric term $P(\Theta)$ under the assumption that there is no lensing (i.e with $\mu=1$) and with the condition that only those events with $SNR>10$ are observed. 
Similarly, to compute the number of lensed events we follow a similar procedure but the SNR is modified by the factor $\sqrt{\mu}>1$ appearing in Eq.~\ref{Eq_ht}. Thanks to the amplifying power of $\mu$, more distant events can make it above the detection threshold provided $\mu$ is large enough to compensate the larger distance $d_L$. 

Once we have computed the number of observed lensed and not lensed events we produce realizations (through Monte Carlo) of these events. First we assign an intrinsic chirp mass and redshift to each event following the probability of having a lensed or not-lensed event for a given combination of (${\rm M}_{c,int},z_{int})$. Then we assign to each event a magnification, $\mu$ (for the lensed events only) and a geometric factor $\Theta$ following the probability distributions of $\mu$ and $\Theta$. Since only events that exceed the detection limit can be observed, both $\mu$ and $\Theta$ determine which events are observable. The role of $\Theta$  is particularly important at high magnification where a double event is expected to be separated by a time delay of $\sim$ several hours. Since  $\Theta$  is largely determined by the relative orientation between the detectors and the line of sight, a separation of a few hours between two identical events originating in the same position in the sky can have significantly different $\Theta$ values so one event may be above the detection limit while the other one falls below.
Most of the observed lensed events will have large magnification factors of order 100 or more. Most of these events will correspond to sources near a fold caustic for which a double image is expected. For perfectly smooth caustics,  the magnification of the two images would be nearly identical. In the presence of microlensing or disruptions to the caustic by susbstructure, the magnifcations of the two images may differ. Microlensing of GW is very small (see however\cite{Lai}) and only when the mass of the microlenses is several thousand solar masses, can substructure induce significant differences between the magnifications (and relative time delay) of the double images. In this work we neglect the role of substructure or microlensing and we assign the same magnification to both events. Hence, only the difference in $\Theta$ determines whether one event is observed while the other one is not. 

To assign a distance to a given event we assume $\Theta=2$ which is a reasonable value for $\Theta$ given the observational bias that events with larger $\Theta$ are more likely to be observed but, on the other hand, large values of $\Theta$ are unlikely. Each simulated event has an associated uncertainty in the distance that is comparable with the uncertainties of the six observed events and that reflects our partial knowledge of $\Theta$. Also, for the lensed events, we estimate their inferred distance which is obtained after assuming the event is interpreted as not being lensed, that is, with $\mu=1$. In this case, an event taking place at a redshift $z_{intr}$ and with magnification $\mu$ will be interpreted as an event taking place at a smaller redshift z such that $D_L(z)=D_L(z_{intr})/\sqrt{\mu}$. If the event had intrinsic chirp mass $M_{c,intr}$ its observed mass is $M_{obs}=M_{c,intr}(1+z_{intr})$ but if it is not interpreted as a lensed event, its intrinsic chirp mass will be erroneously estimated as being $M_{obs}/(1+z)$. That is, it will be biased high by a factor $(1+z_{intr})/(1+z)$.

The distribution of inferred intrinsic chirp masses is then sensitive to the presence or not of lensing.  Lensing increases the fraction of events with high inferred chirp masses as shown in  Fig.~\ref{fig:result4} where we compare the model with Log Normal 1 mass function (see Fig.~\ref{fig:result3}, right panel) plus alternative evolution model (having significant lensing events) with the model where lensing events are rare (broader mass function (Log Normal 2) and a modest evolution model or SFR curve in the left panel of Fig.~\ref{fig:result3}). The two data points are obtained after binning the six observed events in two bins. Clearly, the data favors a very shallow mass function with a very extended tail towards higher masses. Lensing can provide such tail {\emph if} the rate of events at $z\sim 1$--$2$ is sufficiently high.

\end{methods}

\renewcommand{\figurename}{Figure}

\clearpage
\begin{figure}[ht]
\centering
\includegraphics[width=17.5cm]{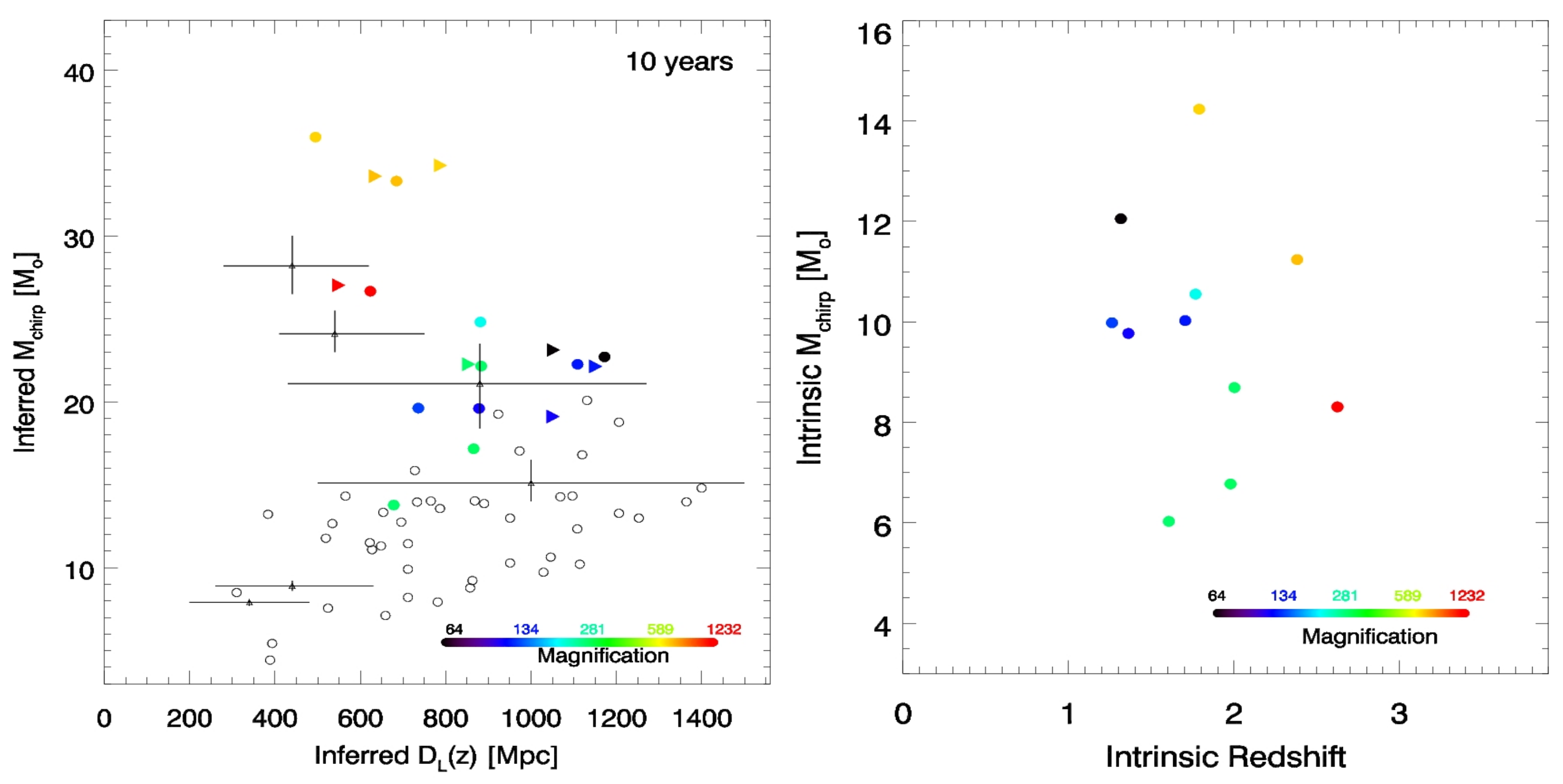}
\caption{{\bf The observed plane of distance and chirp mass:} The six BBH data points with inferred distances and chirp masses and uncertainties reported by the LIGO/Virgo team\cite{Abbot2}. Our predictions for 10 years are shown with unlensed events (open circles) and lensed events coloured by magnification, using a log-normal mass function based on the stellar black holes in our Galaxy, $5<M_\odot<15$ (Log-Normal~1 in Figure~3). The triangle shows a double event (second counterimage) with similar magnification but different geometric factor. Only one event in this simulation had both counterimages with the right visibility configuration to be observed. The redshifts and intrinsic chirp masses of these magnified events reach $z\simeq 2.5$ (right), with evolution as in Figure~3 (Modified SFR) normalised to the lower end of LIGO/Virgo estimates. These lensed events are compared with the data by multiplying the chirp masses by $(1+z)$ and dividing the distances by $\sqrt{\mu}$, as if they were unlensed. The higher chirp mass events appear to define a locus in this plane, of generally increasing magnification with increasing redshift. The predicted unlensed events account well for the location of the two low chirp mass events, but clearly do not reach the highest chirp masses, illustrating the difficulty of accounting for the high chirp masses observed with stellar-mass black holes, without considering lensing.}
\label{fig:result1}
\end{figure}

\clearpage
\begin{figure}[ht]
\centering
\includegraphics[width=17.5cm]{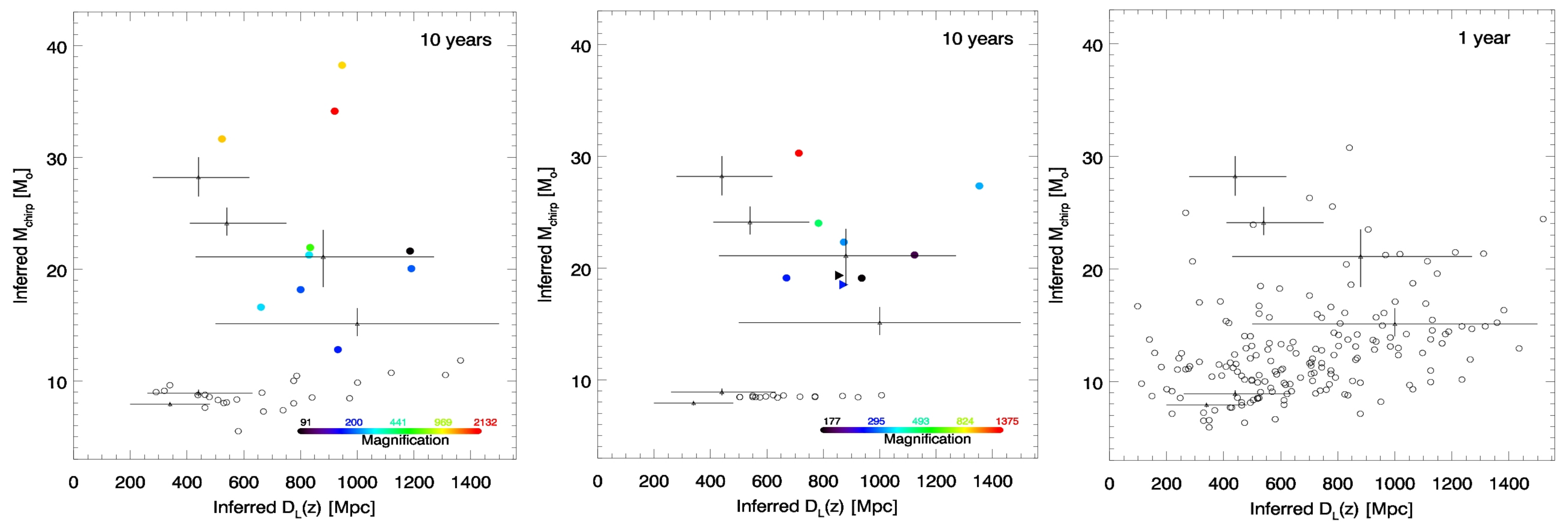}
\caption{{\bf Other models of interest:} We also simulate a purely Gaussian (left) and a monochromatic mass function of $8.5M_\odot$ (middle). As before, triangles show the second counterimage of the same event for those events for which both counterimages would have been detected. For both cases the lensed events fall short of the highest chirp mass events, indicating that a tail of higher black hole masses is preferred, as in Figure~1. On the right, we try to account for the high chirp mass events by drawing the unlensed events from a  broader intrinsic mass function (shown in Figure~3, labelled Log-Normal~2) extending beyond $20M\odot$, and we adopt a normalisation at the high end of the LIGO/Virgo estimates to boost the detection rate of massive unlensed events to help match the data (SFR dashed curve in left panel of Figure~3). This case produces a much smaller proportion of lensed events than the other cases presented (significantly less than one lensed event per year on average). The clear problem here is that within only one year there are far too many events predicted of relatively low chirp mass compared to the data, at odds with distinctive distribution of the data points in this plane.}
\label{fig:result2}
\end{figure}

\clearpage
\begin{figure}[ht]
\centering
\includegraphics[width=17.5cm]{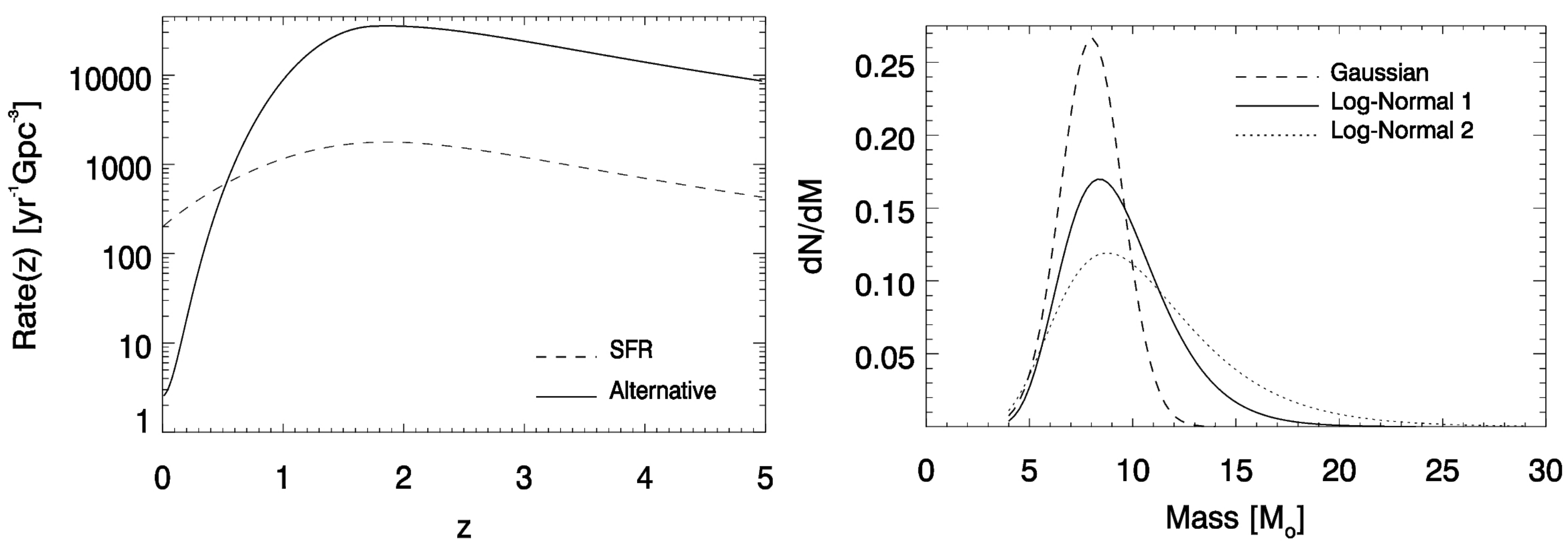}
\caption{{\bf The explored range of BBH event rate evolution and intrinsic chirp mass functions:} On the left we show evolution proportional to the integrated star formation rate (dashed curve) and normalised to the high end of the LIGO/Virgo estimated event rate used in Figure~2 (right). The more steeply declining evolution (solid curve), is preferred by the solution in Figure~1, which is normalised to the lower end of the LIGO/Virgo estimate. On the right, the Gaussian and log-normal mass distributions (see legend) are chosen to be consistent with the known mass range of black hole orbiting stars in our Galaxy. The broader log-normal mass function is used to try account for the  higher chirp mass events in the case of minimal lensing shown in Figure~2 (right).} 
\label{fig:result3}
\end{figure}

\clearpage
\begin{figure}[ht]
\centering
\includegraphics[width=14.0cm]{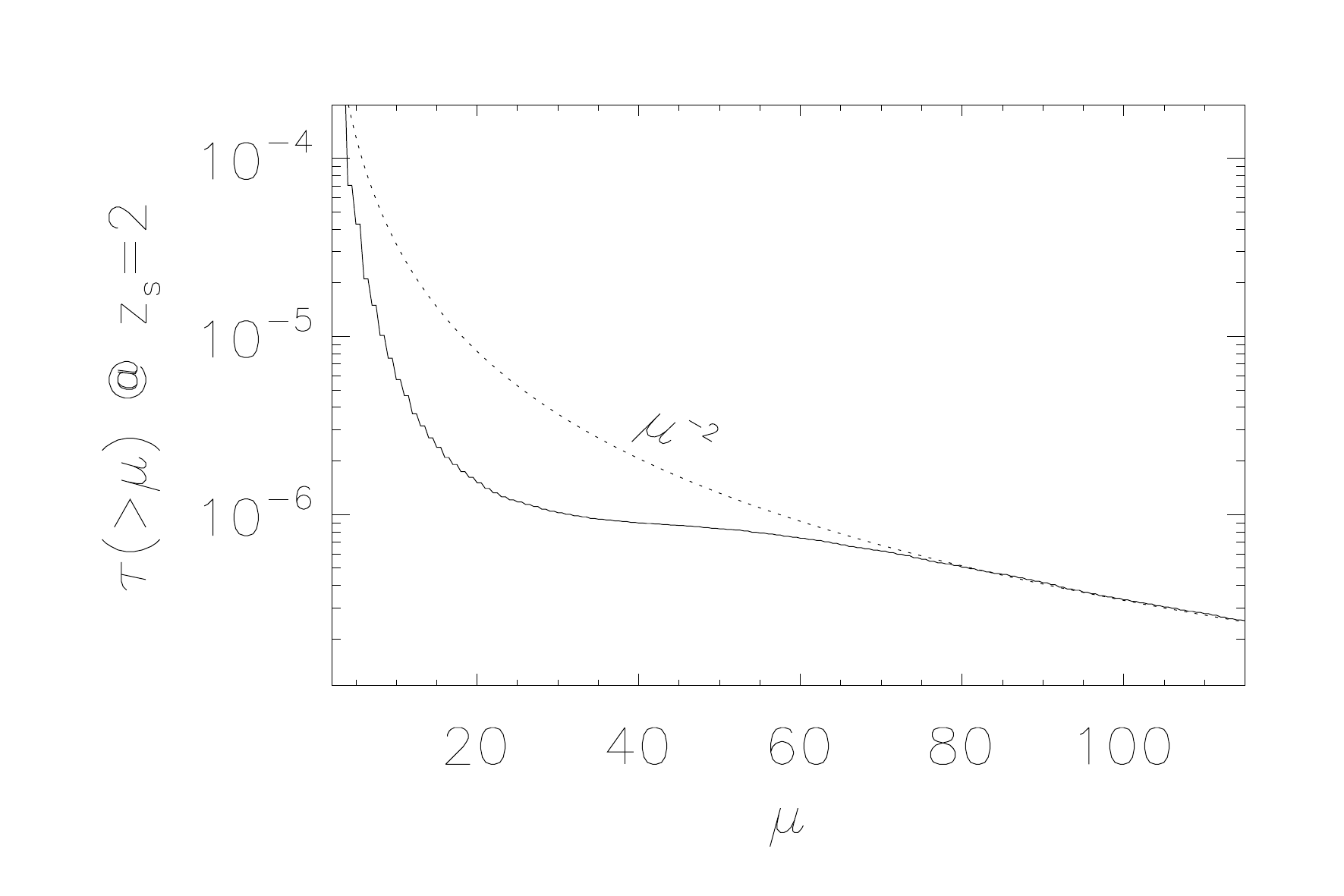}
\caption{{\bf Optical depth for a source at $z_s=2$}. The solid line shows the optical depth for a source at $z_s=2$ and for all lenses with $z<2$. At magnifications larger than $\approx 80$  the probability of having magnification larger than $\mu$ follows the $\mu^{-2}$ law (dotted line), as expected for fold caustics.}
\label{fig:result4}
\end{figure}

\clearpage
\begin{figure}[ht]
\centering
\includegraphics[width=14.0cm]{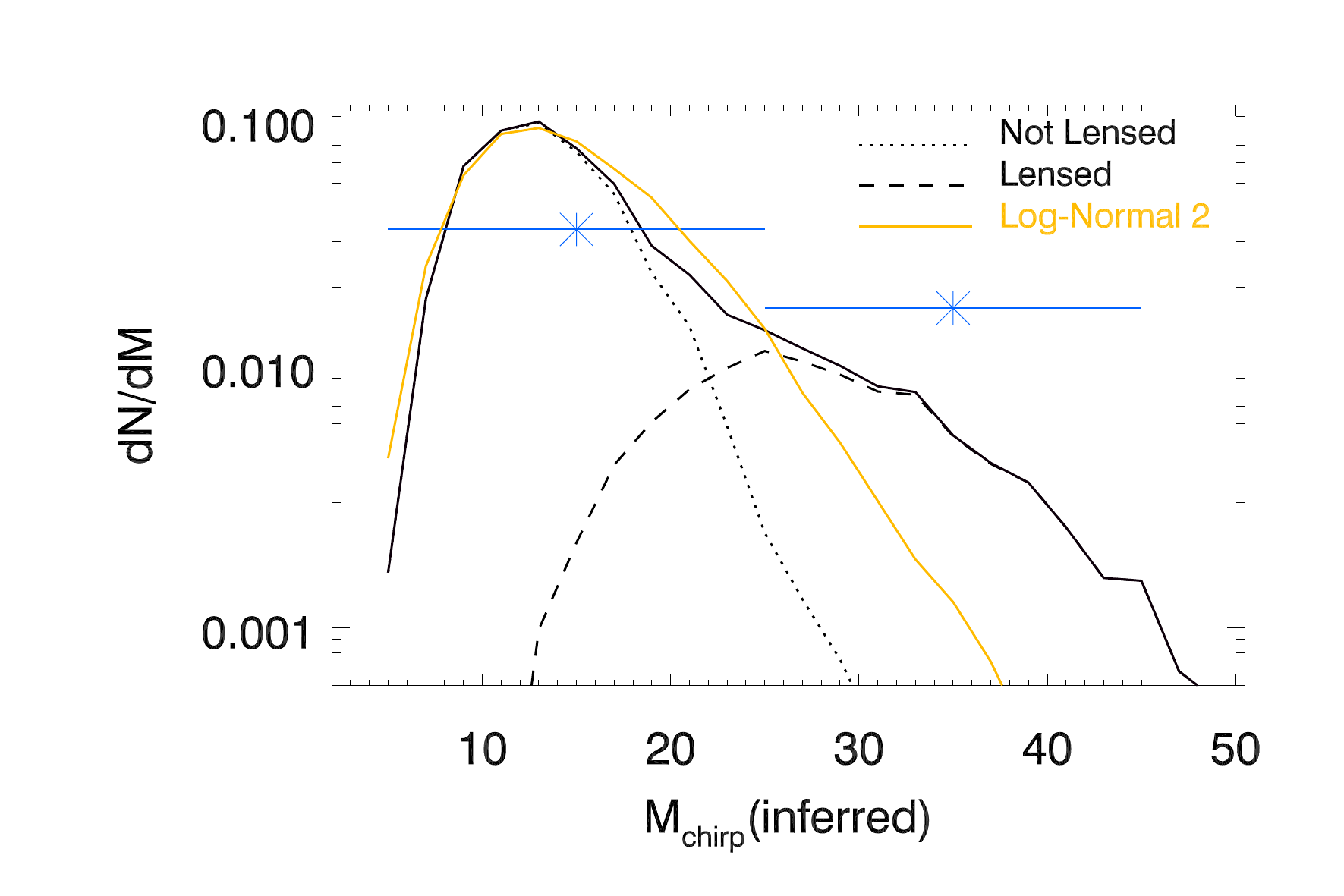}
\caption{{\bf Probability of inferred chirp masses}. The black solid line shows the distribution of observed masses for the proposed model (Log-Normal 1 mass function and alternative evolution model). The lensed and not-lensed events are shown with a dashed and dotted line respectively (the dotted line would correspond to the Log-Normal 1 model without lensing). The color curve is for the model that assumes the Log-Normal2 mass function and the SFR evolution model. In this case, the fraction of lensed events is negligible. The two data points represent the six observed events in the bins $5-25 M_{\odot}$ (4 events) and $25-45 M_{\odot}$ (2 events).} 
\label{fig:result5}
\end{figure}


\clearpage

\begin{thebibliography}{10}
\expandafter\ifx\csname url\endcsname\relax
  \def\url#1{\texttt{#1}}\fi
\expandafter\ifx\csname urlprefix\endcsname\relax\def\urlprefix{URL }\fi
\providecommand{\bibinfo}[2]{#2}
\providecommand{\eprint}[2][]{\url{#2}}






\bibitem{Wang}
\bibinfo{author}{{Wang}, Y.}, \bibinfo{author}{{Stebbins}, A.} \&
  \bibinfo{author}{{Turner}, E.~L.}
\newblock \bibinfo{title}{{Gravitational Lensing of Gravitational Waves from
  Merging Neutron Star Binaries}}.
\newblock \emph{\bibinfo{journal}{Physical Review Letters}}
  \textbf{\bibinfo{volume}{77}}, \bibinfo{pages}{2875--2878}
  (\bibinfo{year}{1996}).
\newblock \eprint{astro-ph/9605140}.

\bibitem{Wardlow}
\bibinfo{author}{{Wardlow}, J.~L.} \emph{et~al.}
\newblock \bibinfo{title}{{HerMES: Candidate Gravitationally Lensed Galaxies
  and Lensing Statistics at Submillimeter Wavelengths}}.
\newblock \emph{\bibinfo{journal}{\apj}} \textbf{\bibinfo{volume}{762}},
  \bibinfo{pages}{59} (\bibinfo{year}{2013}).
\newblock \eprint{1205.3778}.


\bibitem{Bussmann}
\bibinfo{author}{{Bussmann}, R.~S.} \emph{et~al.}
\newblock \bibinfo{title}{{Gravitational Lens Models Based on Submillimeter
  Array Imaging of Herschel-selected Strongly Lensed Sub-millimeter Galaxies at
  z $>$1.5}}.
\newblock \emph{\bibinfo{journal}{\apj}} \textbf{\bibinfo{volume}{779}},
  \bibinfo{pages}{25} (\bibinfo{year}{2013}).
\newblock \eprint{1309.0836}.

\bibitem{Negrello}
\bibinfo{author}{{Negrello}, M.} \emph{et~al.}
\newblock \bibinfo{title}{{The Detection of a Population of
  Submillimeter-Bright, Strongly Lensed Galaxies}}.
\newblock \emph{\bibinfo{journal}{Science}} \textbf{\bibinfo{volume}{330}},
  \bibinfo{pages}{800} (\bibinfo{year}{2010}).
\newblock \eprint{1011.1255}.


\bibitem{Abbot2}
\bibinfo{author}{{Abbott}, B.~P.} \emph{et~al.}
\newblock \bibinfo{title}{{Binary Black Hole Mergers in the First Advanced LIGO
  Observing Run}}.
\newblock \emph{\bibinfo{journal}{Physical Review X}}
  \textbf{\bibinfo{volume}{6}}, \bibinfo{pages}{041015} (\bibinfo{year}{2016}).
\newblock \eprint{1606.04856}.



\bibitem{Abbot1}
\bibinfo{author}{{Abbott}, B.~P.} \emph{et~al.}
\newblock \bibinfo{title}{{Observation of Gravitational Waves from a Binary
  Black Hole Merger}}.
\newblock \emph{\bibinfo{journal}{Physical Review Letters}}
  \textbf{\bibinfo{volume}{116}}, \bibinfo{pages}{061102}
  (\bibinfo{year}{2016}).
\newblock \eprint{1602.03837}.


\bibitem{Remillard}
\bibinfo{author}{{Remillard}, R.~A.}\& \bibinfo{author}{{McClintock}, J.~E.}
\newblock \bibinfo{title}{{X-Ray Properties of Black-Hole Binaries}}.
\newblock \emph{\bibinfo{journal}{\araa}} \textbf{\bibinfo{volume}{44}},
  \bibinfo{pages}{49--92} (\bibinfo{year}{2006}).
\newblock \eprint{astro-ph/0606352}.


\bibitem{Dominik}
\bibinfo{author}{{Dominik}, M.} \emph{et~al.}
\newblock \bibinfo{title}{{Double Compact Objects. II. Cosmological Merger
  Rates}}.
\newblock \emph{\bibinfo{journal}{\apj}} \textbf{\bibinfo{volume}{779}},
  \bibinfo{pages}{72} (\bibinfo{year}{2013}).
\newblock \eprint{1308.1546}.



\bibitem{Belczynski}
\bibinfo{author}{{Belczynski}, K.} \emph{et~al.}
\newblock \bibinfo{title}{{The Effect of Metallicity on the Detection Prospects
  for Gravitational Waves}}.
\newblock \emph{\bibinfo{journal}{\apjl}} \textbf{\bibinfo{volume}{715}},
  \bibinfo{pages}{L138--L141} (\bibinfo{year}{2010}).
\newblock \eprint{1004.0386}.


\bibitem{Abbot3}
\bibinfo{author}{{Abbott}, B.~P.} \emph{et~al.}
\newblock \bibinfo{title}{{Astrophysical Implications of the Binary Black-hole
  Merger GW150914}}.
\newblock \emph{\bibinfo{journal}{\apjl}} \textbf{\bibinfo{volume}{818}},
  \bibinfo{pages}{L22} (\bibinfo{year}{2016}).
\newblock \eprint{1602.03846}.

\bibitem{Bird}
\bibinfo{author}{{Bird}, S.} \emph{et~al.}
\newblock \bibinfo{title}{{Did LIGO Detect Dark Matter?}}
\newblock \emph{\bibinfo{journal}{Physical Review Letters}}
  \textbf{\bibinfo{volume}{116}}, \bibinfo{pages}{201301}
  (\bibinfo{year}{2016}).
\newblock \eprint{1603.00464}.

\bibitem{Carr}
\bibinfo{author}{{Carr}, B.}, \bibinfo{author}{{K{\"u}hnel}, F.} \&
  \bibinfo{author}{{Sandstad}, M.}
\newblock \bibinfo{title}{{Primordial black holes as dark matter}}.
\newblock \emph{\bibinfo{journal}{\prd}} \textbf{\bibinfo{volume}{94}},
  \bibinfo{pages}{083504} (\bibinfo{year}{2016}).
\newblock \eprint{1607.06077}.

\bibitem{Kelly}
\bibinfo{author}{{Kelly}, P.~L.} \emph{et~al.}
\newblock \bibinfo{title}{{An individual star at redshift 1.5 extremely
  magnified by a galaxy-cluster lens}}.
\newblock \emph{\bibinfo{journal}{ArXiv e-prints}}  (\bibinfo{year}{2017}).
\newblock \eprint{1706.10279}.

\bibitem{Venumadhav}
\bibinfo{author}{{Venumadhav}, T.}, \bibinfo{author}{{Dai}, L.} \&
  \bibinfo{author}{{Miralda-Escud{\'e}}, J.}
\newblock \bibinfo{title}{{Microlensing of Extremely Magnified Stars near
  Caustics of Galaxy Clusters}}.
\newblock \emph{\bibinfo{journal}{\apj}} \textbf{\bibinfo{volume}{850}},
  \bibinfo{pages}{49} (\bibinfo{year}{2017}).
\newblock \eprint{1707.00003}.

\bibitem{Diego}
\bibinfo{author}{{Diego}, J.~M.} \emph{et~al.}
\newblock \bibinfo{title}{{Dark matter under the microscope: Constraining
  compact dark matter with caustic crossing events}}.
\newblock \emph{\bibinfo{journal}{ArXiv e-prints}}  (\bibinfo{year}{2017}).
\newblock \eprint{1706.10281}.

\bibitem{Oguri}
\bibinfo{author}{{Oguri}, M.}, \bibinfo{author}{{Diego}, J.~M.},
  \bibinfo{author}{{Kaiser}, N.}, \bibinfo{author}{{Kelly}, P.~L.} \&
  \bibinfo{author}{{Broadhurst}, T.}
\newblock \bibinfo{title}{{Understanding caustic crossings in giant arcs:
  characteristic scales, event rates, and constraints on compact dark matter}}.
\newblock \emph{\bibinfo{journal}{ArXiv e-prints}}  (\bibinfo{year}{2017}).
\newblock \eprint{1710.00148}.

\bibitem{Brandt}
\bibinfo{author}{{Brandt}, T.~D.}
\newblock \bibinfo{title}{{Constraints on MACHO Dark Matter from Compact
  Stellar Systems in Ultra-faint Dwarf Galaxies}}.
\newblock \emph{\bibinfo{journal}{\apjl}} \textbf{\bibinfo{volume}{824}},
  \bibinfo{pages}{L31} (\bibinfo{year}{2016}).
\newblock \eprint{1605.03665}.

\bibitem{Koushiappas}
\bibinfo{author}{{Koushiappas}, S.~M.} \& \bibinfo{author}{{Loeb}, A.}
\newblock \bibinfo{title}{{Dynamics of Dwarf Galaxies Disfavor Stellar-Mass
  Black Holes as Dark Matter}}.
\newblock \emph{\bibinfo{journal}{Physical Review Letters}}
  \textbf{\bibinfo{volume}{119}}, \bibinfo{pages}{041102}
  (\bibinfo{year}{2017}).
\newblock \eprint{1704.01668}.

\bibitem{Green}
\bibinfo{author}{{Green}, A.~M.}
\newblock \bibinfo{title}{{Astrophysical uncertainties on stellar microlensing
  constraints on multisolar mass primordial black hole dark matter}}.
\newblock \emph{\bibinfo{journal}{\prd}} \textbf{\bibinfo{volume}{96}},
  \bibinfo{pages}{043020} (\bibinfo{year}{2017}).
\newblock \eprint{1705.10818}.

\bibitem{Ali-Hamoud}
\bibinfo{author}{{Ali-Ha{\"i}moud}, Y.} \& \bibinfo{author}{{Kamionkowski}, M.}
\newblock \bibinfo{title}{{Cosmic microwave background limits on accreting
  primordial black holes}}.
\newblock \emph{\bibinfo{journal}{\prd}} \textbf{\bibinfo{volume}{95}},
  \bibinfo{pages}{043534} (\bibinfo{year}{2017}).
\newblock \eprint{1612.05644}.

\bibitem{Gaggero}
\bibinfo{author}{{Gaggero}, D.} \emph{et~al.}
\newblock \bibinfo{title}{{Searching for Primordial Black Holes in the Radio
  and X-Ray Sky}}.
\newblock \emph{\bibinfo{journal}{Physical Review Letters}}
  \textbf{\bibinfo{volume}{118}}, \bibinfo{pages}{241101}
  (\bibinfo{year}{2017}).
\newblock \eprint{1612.00457}.



\bibitem{Finn}
\bibinfo{author}{{Finn}, L.~S.}
\newblock \bibinfo{title}{{Binary inspiral, gravitational radiation, and
  cosmology}}.
\newblock \emph{\bibinfo{journal}{\prd}} \textbf{\bibinfo{volume}{53}},
  \bibinfo{pages}{2878--2894} (\bibinfo{year}{1996}).
\newblock \eprint{gr-qc/9601048}.


\bibitem{Ng}
\bibinfo{author}{Ng, K. K.~Y.}, \bibinfo{author}{Wong, K. W.~K.},
  \bibinfo{author}{Li, T. G.~F.} \& \bibinfo{author}{Broadhurst, T.}
\newblock \bibinfo{title}{{Precise LIGO Lensing Rate Predictions for Binary
  Black Holes}}.
\newblock \emph{\bibinfo{journal}{Phys. Rev.}} \textbf{\bibinfo{volume}{B96}},
  \bibinfo{pages}{155205} (\bibinfo{year}{2017}).
\newblock \eprint{1703.06319}.


\bibitem{Turner}
\bibinfo{author}{{Turner}, E.~L.}, \bibinfo{author}{{Ostriker}, J.~P.} \&
  \bibinfo{author}{{Gott}, J.~R., III}.
\newblock \bibinfo{title}{{The statistics of gravitational lenses - The
  distributions of image angular separations and lens redshifts}}.
\newblock \emph{\bibinfo{journal}{\apj}} \textbf{\bibinfo{volume}{284}},
  \bibinfo{pages}{1--22} (\bibinfo{year}{1984}).


\bibitem{Madau}
\bibinfo{author}{{Madau}, P.} \& \bibinfo{author}{{Dickinson}, M.}
\newblock \bibinfo{title}{{Cosmic Star-Formation History}}.
\newblock \emph{\bibinfo{journal}{\araa}} \textbf{\bibinfo{volume}{52}},
  \bibinfo{pages}{415--486} (\bibinfo{year}{2014}).
\newblock \eprint{1403.0007}.


\bibitem{Broadhurst}
\bibinfo{author}{{Broadhurst}, T.} \& \bibinfo{author}{{Lehar}, J.}
\newblock \bibinfo{title}{{A Gravitational Lens Solution for the IRAS Galaxy
  FSC 10214+4724}}.
\newblock \emph{\bibinfo{journal}{\apjl}} \textbf{\bibinfo{volume}{450}},
  \bibinfo{pages}{L41} (\bibinfo{year}{1995}).
\newblock \eprint{astro-ph/9505013}.

\bibitem{Schmidt}
\bibinfo{author}{{Schmidt}, M.}
\newblock \bibinfo{title}{{Statistical Studies of the Evolution of
  Extragalactic Radio Sources. I. Quasars.}}
\newblock \emph{\bibinfo{journal}{\apj}} \textbf{\bibinfo{volume}{176}},
  \bibinfo{pages}{273} (\bibinfo{year}{1972}).

\bibitem{Smith}
\bibinfo{author}{{Smith}, G.~P.} \emph{et~al.}
\newblock \bibinfo{title}{{What if LIGO's gravitational wave detections are
  strongly lensed by massive galaxy clusters?}}
\newblock \emph{\bibinfo{journal}{\mnras}}  (\bibinfo{year}{2018}).
\newblock \eprint{1707.03412}.



\bibitem{Nakamura}
\bibinfo{author}{{Nakamura}, T.~T.}
\newblock \bibinfo{title}{{Gravitational Lensing of Gravitational Waves from
  Inspiraling Binaries by a Point Mass Lens}}.
\newblock \emph{\bibinfo{journal}{Physical Review Letters}}
  \textbf{\bibinfo{volume}{80}}, \bibinfo{pages}{1138--1141}
  (\bibinfo{year}{1998}).




\bibitem{Hilbert}
\bibinfo{author}{{Hilbert}, S.}, \bibinfo{author}{{White}, S.~D.~M.},
  \bibinfo{author}{{Hartlap}, J.} \& \bibinfo{author}{{Schneider}, P.}
\newblock \bibinfo{title}{{Strong-lensing optical depths in a {$\Lambda$}CDM
  universe - II. The influence of the stellar mass in galaxies}}.
\newblock \emph{\bibinfo{journal}{\mnras}} \textbf{\bibinfo{volume}{386}},
  \bibinfo{pages}{1845--1854} (\bibinfo{year}{2008}).
\newblock \eprint{0712.1593}.

\bibitem{Zackrisson}
\bibinfo{author}{{Zackrisson}, E.} \emph{et~al.}
\newblock \bibinfo{title}{{Primordial star clusters at extreme magnification}}.
\newblock \emph{\bibinfo{journal}{\mnras}} \textbf{\bibinfo{volume}{449}},
  \bibinfo{pages}{3057--3063} (\bibinfo{year}{2015}).
\newblock \eprint{1411.6628}.

\bibitem{Dai1}
\bibinfo{author}{{Dai}, L.}, \bibinfo{author}{{Venumadhav}, T.} \&
  \bibinfo{author}{{Sigurdson}, K.}
\newblock \bibinfo{title}{{Effect of lensing magnification on the apparent
  distribution of black hole mergers}}.
\newblock \emph{\bibinfo{journal}{\prd}} \textbf{\bibinfo{volume}{95}},
  \bibinfo{pages}{044011} (\bibinfo{year}{2017}).
\newblock \eprint{1605.09398}.



\bibitem{Dai2}
\bibinfo{author}{{Dai}, L.} \& \bibinfo{author}{{Venumadhav}, T.}
\newblock \bibinfo{title}{{On the waveforms of gravitationally lensed
  gravitational waves}}.
\newblock \emph{\bibinfo{journal}{ArXiv e-prints}}  (\bibinfo{year}{2017}).
\newblock \eprint{1702.04724}.

\bibitem{Blandford}
\bibinfo{author}{{Blandford}, R.} \& \bibinfo{author}{{Narayan}, R.}
\newblock \bibinfo{title}{{Fermat's principle, caustics, and the classification
  of gravitational lens images}}.
\newblock \emph{\bibinfo{journal}{\apj}} \textbf{\bibinfo{volume}{310}},
  \bibinfo{pages}{568--582} (\bibinfo{year}{1986}).
  
  
   
\bibitem{Watson13}
\bibinfo{author}{{Watson}, W.~A.} , \bibinfo{author}{{Iliev}, I.~T.} \emph{et~al.}
\newblock \bibinfo{title}{{The halo mass function through the cosmic ages}}.
\newblock \emph{\bibinfo{journal}{\mnras}} \textbf{\bibinfo{volume}{433}},
  \bibinfo{pages}{1230--1245} (\bibinfo{year}{2013}).
  
   
\bibitem{Watson14}
\bibinfo{author}{{Watson}, W.~A.}, \bibinfo{author}{{Diego}, J.~M.} \emph{et~al.}
\newblock \bibinfo{title}{{The Jubilee ISW project - I. Simulated ISW and weak lensing maps and initial power spectra results}}.
\newblock \emph{\bibinfo{journal}{\mnras}} \textbf{\bibinfo{volume}{438}},
  \bibinfo{pages}{412--425} (\bibinfo{year}{2014}).
  
  
  
\bibitem{Prada}
\bibinfo{author}{{Prada}, F.}  \emph{et~al.}
\newblock \bibinfo{title}{{Halo concentrations in the standard Λ cold dark matter cosmology}}.
\newblock \emph{\bibinfo{journal}{\mnras}} \textbf{\bibinfo{volume}{423}},
  \bibinfo{pages}{3018--3030} (\bibinfo{year}{2012}).
  
 
\bibitem{Lai}
\bibinfo{author}{{Lai}, K.~H.}  \emph{et~al.}
\newblock \bibinfo{title}{{Discovering intermediate-mass black hole lenses through gravitational wave lensing}}.
\newblock \emph{\bibinfo{journal}{\prd}} \textbf{\bibinfo{volume}{--}},
  \bibinfo{pages}{--} (\bibinfo{year}{2018}).
  
  
  

\end{thebibliography}
\end{document}